\pdfoutput=1
\documentclass[twocolumn,10pt,a4paper,floatfix,prl,unsortedaddress,superscriptaddress,longbibliography, nofootinbib]{revtex4-2}

\usepackage[utf8]{inputenc}
\usepackage{microtype}
\usepackage{newtxtext}
\usepackage[upint]{newtxmath}
\usepackage{eucal}

\usepackage{comment}
\usepackage{graphicx}
\usepackage{booktabs}
\usepackage{enumerate}
\usepackage{siunitx}
\usepackage{braket}
\usepackage{color}
\usepackage{float}
\usepackage{soul}
\usepackage{bm}

\usepackage[colorlinks,allcolors=blue]{hyperref}
\usepackage[capitalize]{cleveref}

\usepackage[svgnames]{xcolor}
\usepackage{todonotes}
\setuptodonotes{inline}

\definecolor{DarkRed}{rgb}{0.80,0,0}
\definecolor{DarkBlue}{rgb}{0.20,0,0.80}
\definecolor{Purple}{rgb}{0.55,0,0.55}
\definecolor{Magenta}{rgb}{1,0,1}

\begin{document}
\title{Hidden Spin-Supercurrent Counterflow in Conventional Superconductors}
\author{Johanne Bratland Tjernshaugen}
\email[Contact author: ]{johanne.b.tjernshaugen@ntnu.no}
\affiliation{Center for Quantum Spintronics, Department of Physics, Norwegian \\ University of Science and Technology, NO-7491 Trondheim, Norway}
\author{Jacob Linder}
\affiliation{Center for Quantum Spintronics, Department of Physics, Norwegian \\ University of Science and Technology, NO-7491 Trondheim, Norway}

\begin{abstract}
\noindent 
Resistive spin currents conventionally vary in space through spin relaxation or torque, which removes angular momentum from the transported spin. We here demonstrate a qualitatively different mechanism in a conventional superconductor exposed to a spin-splitting field. Spin injection across a spin-mixing interface activates a spin-supercurrent contribution, allowing the resistive spin current to vary spatially while the total spin current remains conserved. Even without a net spin-bias across the superconductor, finite and oppositely directed resistive and condensate spin currents arise. This produces a hidden spin-supercurrent counterflow with vanishing total spin and charge currents. The mechanism persists away from exact compensation and remains sizable over broad parameter ranges. The counterflow may be accessible through spin-to-charge conversion, either as a modification of the electrical response of spin-Hall injectors or through a local quasiparticle inverse spin-Hall signal. Our results establish condensate-mediated spin-current compensation as a previously overlooked alternative to spin relaxation.
\end{abstract}

\maketitle

\textit{Introduction}--- 
Understanding the decay and conservation of spin currents in various materials is essential in spintronics. In normal materials, spin currents decay due to magnetic impurities~\cite{watts2019magnetic}, spin-orbit scattering~\cite{elliott1954theory, yafet1963, zutic2004spintronics, fabian1999phonon}, or the transfer of spin angular momentum caused by the torque from a magnetization \cite{ralph2008spin}. In either case, a decrease in the spin current signals that angular momentum has been transferred out of the transported spin.

In conventional superconductors, injected spin is transported by quasiparticles alone since the spin-singlet Cooper pairs carry no spin~\cite{morten2008proximity}. The opening of the superconducting gap renormalizes the relaxation rate of the quasiparticle spins~\cite{yafet1983conduction,yang_natmat_10, silaev2015spin, morten2004spin, morten2005spin, poli2008spin}. Of particular interest is spin injection into spin-split superconductors where spin couples to energy transport~\cite{bergeret_rmp_18, beckmann2024spin} and where long spin-diffusion lengths have been observed~\cite{ hubler_prl_12, quay2013spin}.  Such nonlocal spin-transport experiments typically interpret spatial decay of the quasiparticle spin signal as spin relaxation.

However, a superconductor can support additional channels for angular-momentum transfer. In particular, mobile Abrikosov vortices were recently shown to provide a dissipative angular-momentum channel near $T_c$~\cite{mangold2026spin}. More generally, the superconducting condensate can provide a second transport channel for spin in the presence of triplet Cooper pairs. Such Cooper pairs can either be intrinsic or proximity-induced via magnetic materials~\cite{bergeret_rmp_05, buzdin_rmp_05, eschrig_rpp_15, linder_natphys_15}. Previous works have shown that injected spin can be transferred to the condensate in intrinsic triplet superconductors~\cite{li2026pumping}, and that proximity-induced equal-spin Cooper pairs can participate in nonequilibrium spin transport  \cite{jeon_natmat_18, ojajarvi_prl_22, montiel2023spin}. It has also been shown that equilibrium spin supercurrents are conserved when injected into conventional superconductors, even in the presence of magnetic impurities and spin-orbit scattering~\cite{ouassou2017conservation}. A uniform spin accumulation acting on such an equilibrium spin supercurrent can generate a transverse component of the spin flow  through a torque-like coupling \cite{ouassou2019controlling}.

The above establishes that quasiparticles and the condensate can both contribute to a net spin flow, suggesting that such currents only exist in the presence of a net spin current. Here, we show that superconductivity provides a qualitatively different mechanism for spin flow where this is not the case. Rather than being relaxed or transferred to a magnetization, the quasiparticle spin current varies spatially because it is locally compensated by an oppositely directed spin supercurrent. The two contributions form a hidden counterflow, allowing the quasiparticle current to vary in space while the total spin current remains zero. We find that this compensation mechanism also persists for a finite conserved spin flow.

To illustrate the counterflow, we consider a conventional $s$-wave superconductor with a spin splitting field, and place it in contact with a spin accumulation reservoir across spin-mixing interfaces. We show that both the quasiparticle current and the spin supercurrent vary spatially in the superconductor, even in the absence of spin relaxation, but their contributions are hidden inside a conserved net spin current. This exchange of angular momentum between quasiparticles and the equal-spin Cooper pairs affects how the spatial variation of quasiparticle spin currents in spin-transport experiments should be interpreted. 
Furthermore, we show that the counterflow remains very robust against variations in the junction parameters. We propose that an experimental signature of this counterflow is accessible through spin-to-charge conversion, and estimate an inverse spin-Hall voltage of order 10 nanovolts.

\begin{figure}
    \centering
    \includegraphics[]{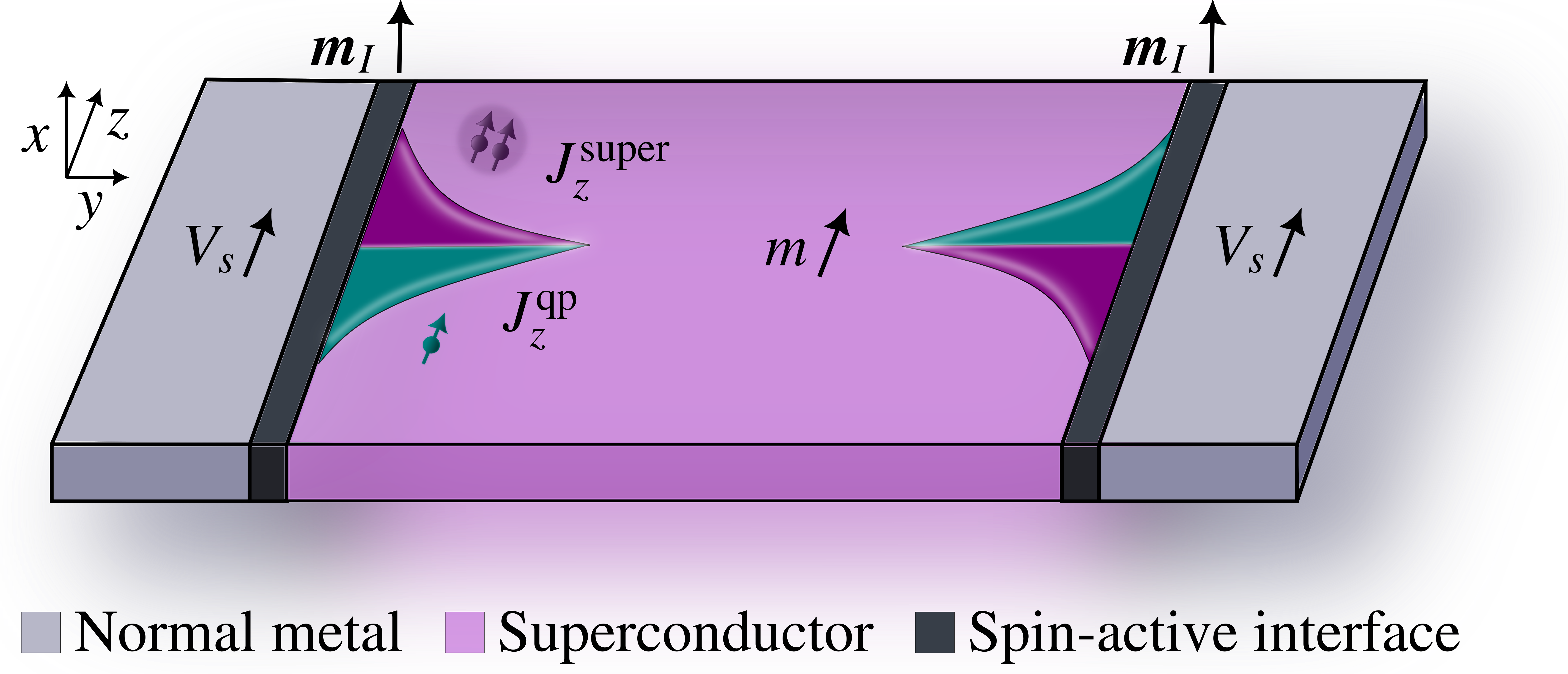}    
    \caption{A singlet superconductor with in-plane spin splitting $m$ along $z$ is connected to two normal reservoirs. The normal reservoirs have spin accumulations ${V}_s$ polarized along $z$. The interfaces are tunneling with spin mixing, and the interface magnetization $\boldsymbol{m}_I$ is orthogonal to the spin accumulation. It is not necessary to have the spin splitting field perfectly aligned with the spin accumulation, and it is also not necessary to have $\boldsymbol{m}_I$ perfectly orthogonal to it.  The net spin current is zero both above and below the critical temperature of the superconductor. However, below $T_c$, it consists of two non-zero, but cancelling currents: the resistive spin current $J_z^{\text{qp}}$, carried by quasiparticles, and the spin supercurrent $J_z^{\text{super}}$, carried by triplet Cooper pairs. This causes a spatial variation of the resistive spin current, appearing exclusively below $T_c$ and acting as a signature of the counterflowing spin supercurrent. 
    } 
    \label{fig:setup}
\end{figure}

\textit{Methodology}--- We model the system shown in Fig.~\ref{fig:setup} using self-consistent non-equilibrium Keldysh-Usadel Green's function theory. The isotropic, quasiclassical Green's function in Keldysh$\otimes$Nambu$\otimes$spin space is given by 
\begin{equation}
    \check{g}=\begin{pmatrix}
        \hat{g}^R & \hat{g}^K \\ 0 & \hat{g}^A
    \end{pmatrix},
\end{equation}
where $\hat{g}^R$ is the retarded Green's function, $\hat{g}^A=-\hat{\rho}_4(\hat{g}^R)^\dagger \hat{\rho}_4$ is the advanced Green's function, $\hat{g}^K=\hat{g}^R\hat{h}-\hat{h}\hat{g}^A$ is the Keldysh Green's function, and $\hat{h}=h_n\hat{\rho}_n$ is the distribution function. The matrices $\hat{\rho}_n$ span the block-diagonal Nambu$\otimes$spin space, and they are defined as $\hat{\rho}_n = \hat{\tau}_0 \hat{\sigma}_n$ and $\hat{\rho}_{n+4} = \hat{\tau}_3 \hat{\sigma}_{n}$ for $n\in \{0,1,2,3\}$ with $\hat{\tau}_0=\mathrm{diag}(\sigma_0, \sigma_0)$, $\hat{\tau}_3=\mathrm{diag}(\sigma_0, -\sigma_0)$ and $\hat{\sigma}_n = \mathrm{diag}(\sigma_n, \sigma^*_n)$. The Green's function obeys the Usadel equation, which is valid in the diffusive limit where the mean free path $l_\text{mfp}$ is the shortest length scale apart from the Fermi wavelength:
\begin{equation}
    \xi^2{\frac{\partial}{\partial y}\left(\check{g} \frac{\partial\check{g}}{\partial y}\right) }= -i[\varepsilon\hat{\rho}_4+\hat{M}+\hat{\Delta},\check{g}]/\Delta_0.
\end{equation}
Here, $\xi=\sqrt{\hbar D/\Delta_0}$ is the superconducting coherence length, $D$ is the electron diffusion coefficient of the normal-state metal, $\varepsilon$ is the energy, and $\Delta_0$ is the bulk superconducting gap at zero temperature. The spin-splitting of strength $m$ in the $z$-direction is included via the self-energy $\hat{M} = m \hat{\sigma}_z
$. The self-energy $\hat{\Delta} = \text{antidiag}(\Delta, -\Delta, \Delta^*, -\Delta^*)$ is the self-consistent superconducting order parameter. Inelastic scattering is modeled using the Dynes approximation $\varepsilon \rightarrow \varepsilon + i\delta$ in the retarded equation. The normal metals are modelled as reservoirs with $\hat{g}_N^R=\hat{\rho}_4$, and the spin accumulation $V_s$ enters via the distribution function
\begin{equation}
    \hat{h}_N =\frac{1}{2}(t_++t_-)\hat{\rho}_0 + \frac{1}{2}(t_+ - t_-)\hat{\rho}_7  
\end{equation}
where $t_\pm = \text{tanh}[(\varepsilon\pm V_s)/2T]$ and $T$ is the temperature. The superconducting region is coupled to the normal reservoirs with spin-active boundary conditions,
\begin{equation}
    \check{g}\frac{\partial}{\partial(y/\xi)}\check{g} = 
    \frac{\pm 1}{2\chi}
    \left[\check{g}_N  -i \frac{G_{\varphi}}{G_0}\check{m}_I , \check{g}\right].
\end{equation}
Here, $\pm$ is valid for $y=0$ and $y=l$, respectively, and $l$ is the junction length. 
The interface parameter $\chi = A\sigma/G_0\xi$
depends on the interface cross-sectional area $A$, the bulk normal-state conductivity $\sigma$, 
and the tunneling conductance $G_0$. $G_\varphi$ is the spin-mixing, and the average interface magnetization felt by a reflected quasiparticle at the interface is described by $\check{m}_I = \hat{1}\otimes \text{diag}(\boldsymbol{m}_I\cdot \boldsymbol{\sigma}, \boldsymbol{m}_I\cdot \boldsymbol{\sigma}^*)$. We assume that the interface polarization is zero, since such a polarization is not required for the counterflow predicted here, meaning even weakly magnetic interfaces are sufficient.


The spin current is carried by quasiparticles and triplet Cooper pairs, and is therefore a sum of the resistive quasiparticle current and the spin supercurrent: $J_\nu = J_\nu^\mathrm{qp}+J_\nu^\mathrm{super}$. The resistive spin current originates from the spatial variation of the distribution function, and is therefore the part of $J_\nu$ that depends on $\partial_y\hat{h}$. The spin supercurrent is the part that depends $\hat{h}$, but not its derivative. The latter contribution originates from the superconducting spectral current and is weighted by the local distribution function without requiring a distribution gradient. It can therefore remain finite for a spatially uniform distribution and vanishes when the superconducting spectral correlations are removed, identifying it as the spin-supercurrent contribution rather than a diffusion-driven quasiparticle current. In the presence of a pure spin accumulation, the nonzero components of the distribution function are $h_0$ and $\boldsymbol{h}_s=(h_5,h_6,h_7)=(h_x,h_y,h_z)$ and the spin supercurrent can be simplified to \cite{ouassou2019controlling}
\begin{equation}\label{eq:spin supercurrent}
    J_\nu^\mathrm{super} = - J_0 \int_{0}^\infty \mathrm{d}\left(\frac{\varepsilon}{\Delta_0}\right) [\mathrm{Im}(\boldsymbol{A}) h_0 + \mathrm{Re}(\boldsymbol{A})\times \boldsymbol{h}_s]_\nu,
\end{equation}
where $J_0=N_0\Delta_0^2\xi/2$, $N_0$ is the normal-state density of states at $T=0$  and $\boldsymbol{A} = \xi(\boldsymbol{g}_t\times\partial_{y} \boldsymbol{g}_t - \boldsymbol{f}_t\times \partial_{y}\Tilde{\boldsymbol{f}}_t)$ and the retarded Green's function was parametrized as 
\begin{equation}
     \hat{g}^R = \begin{pmatrix}
        g_s+\boldsymbol{g}_t\cdot \boldsymbol{\sigma} & (f_s+\boldsymbol{f}_t\cdot \boldsymbol{\sigma})i\sigma_2\\ -i\sigma_2(\Tilde{f}_s - \Tilde{\boldsymbol{f}}_t\cdot\boldsymbol{\sigma}) & -\sigma_2(\Tilde{g}_s-\tilde{\boldsymbol{g}}_t\cdot \boldsymbol{\sigma})\sigma_2
    \end{pmatrix}.
\end{equation}
The charge and spin-energy modes $h_i\; ( i=1,2,3,4)$ are all zero in the present system. The resistive spin current driven by gradients in the distribution functions is \cite{ouassou2018voltage}:
\begin{equation}\label{eq:qp spincurrent}
     J^\text{qp}_\nu = J_0 \int_0^\infty \text{d}\Big(\frac{\varepsilon}{\Delta_0}\Big)M_{\nu 0}\partial_{y/\xi}h_0 + \boldsymbol{M}_\nu \cdot \partial_{y/\xi}\boldsymbol{h}_s 
\end{equation}
where $M_{\nu 0}=\mathrm{Tr}[- \hat{\tau}_3\hat{\sigma}_\nu\hat{g}^R\hat{g}^A]/8$ and $[\boldsymbol{M}_\nu]_{\nu'} = \mathrm{Tr}[\hat{\sigma}_\nu\hat{\sigma}_{\nu'} - \hat{\tau}_3\hat{\sigma}_\nu\hat{g}^R\hat{\tau}_3\hat{\sigma}_{\nu'}\hat{g}^A]/8$ are the spectral transport coefficients. In the spin-degenerate normal state limit, this reduces to $J_\nu^{\mathrm{qp}} \propto \int^\infty_0 d\varepsilon \partial_y h_\nu$, analogous to Fick's law for diffusion. 

In the absence of spin relaxation, the spin current component parallel to the spin splitting in the superconductor is conserved. Spin current components perpendicular to the spin splitting are affected by bulk torque and will therefore decay. Spin relaxation from magnetic impurities can be disregarded by making $l$ much shorter than the spin relaxation length $l_\text{sf}$, while still making it long enough to stay in the diffusive regime. The required regime $l_\text{mfp} \ll l \ll l_\text{sf}$ is fulfilled \textit{e.g.} in Al \cite{hubler_prl_12, liu_prb_22}.

The setup in Fig. \ref{fig:setup} has a superconductor with in-plane spin splitting, so that no Meissner currents exist, connected to normal reservoirs with spin accumulation polarized along $z$. The interface magnetization $\boldsymbol{m}_I$ is taken to point along $x$, while $y$ is the effective 1D junction coordinate. Importantly, it is sufficient that the spin splitting has some component parallel to the spin accumulation, and that the interface magnetization has some component perpendicular to the spin accumulation. Consequently, the spin counterflow effect predicted here does not require a specially engineered perpendicular configuration or precise fine-tuning of the vector directions, which considerably broadens the range of experimentally accessible realizations.

\begin{figure}
    \centering
    \includegraphics{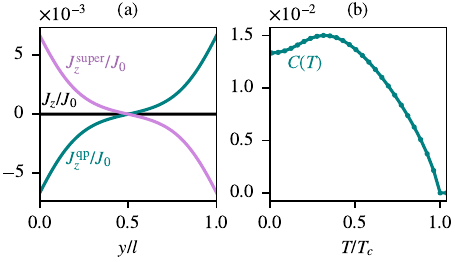}
    \caption{(a) Decomposition of the net-zero $z$-polarized spin current $J_z$ into a quasiparticle contribution $J_z^\text{qp}$ and a condensate contribution $J_z^\text{super}$ at $T/T_c= 0.011$. (b) Counterflow measure vs. temperature. The counterflow measure is defined as $C=[\text{max}_y(J_z^\text{qp})-\text{min}_y(J_z^\text{qp})]/J_0$, describing the peak-to-peak variation of the resistive spin current due to spin supercurrent counterflow. The parameters used are $l=3\xi$, $G_\varphi/G_0=3$, $\chi=9$, $m=0.25\Delta_0$, and $V_s=0.15\Delta_0$.
    }
    \label{fig:mainres}
\end{figure}

\textit{Spin counterflow}--- We first consider the normal state $T>T_c$. When both reservoirs have equal spin accumulation $V_s$, the $z$-polarized spin current $J^\text{qp}_z$ is zero. The combination of the spin splitting $m$ in the superconductor and spin-mixing $G_\varphi$ at the interfaces generates $h_x$ and $h_y$ modes in addition to the injected $h_z$ mode from the reservoirs. However, there is no decay mechanism for $J^\text{qp}_z$ or diffusive driving force since the reservoirs have equal $V_s$. This causes the $z$-polarized spin current to vanish. 

Next, we consider the superconducting state $T<T_c$. Superconducting equal-spin triplet correlations $f_{\uparrow\uparrow}$ and  $f_{\downarrow\downarrow}$ arise which generate a finite spectral component $A_y$, while the reservoir spin accumulation maintains the nonequilibrium spin modes $\boldsymbol{h}_s$ already present from the normal state. 
Since $\boldsymbol{A}\parallel \hat{\boldsymbol{y}}$, $h_x$ induces a $z$-polarized spin supercurrent according to the cross product in Eq. \eqref{eq:spin supercurrent}. In effect, $J_z^\mathrm{super}$ only arises when the system is driven out of equilibrium. However, since both $A_y$ and $h_x$ vary in space for $T<T_c$, $J_z^\mathrm{super}$ varies spatially. Since the total spin current must be conserved when $l \ll l_\text{sf}$, a quasiparticle spin current is forced to arise to balance the spin supercurrent. This is the hidden spin supercurrent counterflow: a net zero spin current, made up of canceling quasiparticle and triplet Cooper pair spin currents, triggered by the superconducting transition $T<T_c$. The currents are shown in Fig. \ref{fig:mainres}(a). The resulting current profiles have the antisymmetric spatial parity required by inversion symmetry. To quantify the magnitude of the counterflow, we define
\begin{equation}
    C=[\text{max}_y(J_z^\text{qp})-\text{min}_y(J_z^\text{qp})]/J_0.
\end{equation}
We show the temperature dependence of $C$ in Fig. \ref{fig:mainres}(b). Above $T_c$, the quasiparticle spin current is conserved and $C=0$. 
Below $T_c$, the noncollinear magnetic structure generates multiple triplet components. Consequently, the full spin-Nambu retarded Green function cannot generally be reduced to independent spin sectors by a single global choice of spin-quantization axis. Its normal and anomalous spin structures vary with energy and position, resulting in off-diagonal spectral transport coefficients such as $M_{zx}$ and $M_{zy}$.
Physically, this means that gradients of the  occupation modes $h_x$ and $h_y$ can drive a $z$-polarized quasiparticle current because the quasiparticles to which they couple can also carry a finite $z$-spin projection. This is reflected in the non-zero spectral transport coefficients ${M}_{zx}$ and ${M}_{zy}$. The resulting $C(T)$ is weakly nonmonotonic and reaches a maximum near $T\simeq0.3T_c$. A decomposition into the four different kinetic-mode contributions, i.e. $M_{z0}\partial_y h_0$ and $\boldsymbol{M}_z\cdot \partial_y \boldsymbol{h}_s$ in Eq. \eqref{eq:qp spincurrent}, shows that the nonmonotonic $C(T)$ behavior results from a redistribution between the channels (see the End Matter). At higher temperatures, the reduction of the superconducting gap suppresses the triplet correlations and the noncollinear spectral structure responsible for both the off-diagonal transport response and the counterflow. All contributions therefore vanish at $T_c$. The emergence of superconductivity thus introduces a qualitatively new mechanism for quasiparticle spin-current decay beyond impurity relaxation and spin-transfer torque. Quasiparticle spin flow can be continuously converted into a counterflowing spin supercurrent, allowing the quasiparticle contribution to decrease while the total spin current remains conserved. Thus, even a conventional BCS superconductor enables a redistribution of spin transport between resistive and dissipationless channels.

\begin{figure}
    \centering
    \includegraphics{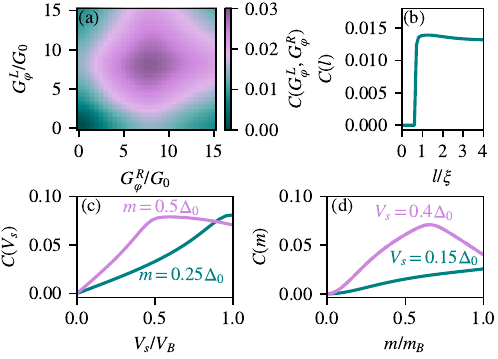}
    \caption{The counterflow remains sizable over broad parameter ranges, while its magnitude can be enhanced by optimizing experimentally tunable parameters. (a) The dependence of $C$ on the spin-mixing $G_{\varphi}^{L/R}$ of the left and right interfaces, respectively. (b)-(d): Variation of $C$ with junction length, spin accumulation, and spin splitting, respectively. The normalization constants $V_B(m)$ and $m_B(V_s)$ denote the maximal values for the spin accumulation and the spin splitting, respectively, before the superconducting state enters a bistable regime. These values are given in the End Matter. 
    The parameters used are the same as in Fig. \ref{fig:mainres}, unless stated otherwise, and the temperature is $T/T_c=0.011$. 
    }
    \label{fig:robustness and maximization}
\end{figure}

\textit{Robustness and maximization}--- The dependence of the spin current counterflow magnitude on the system parameters is shown in Fig. \ref{fig:robustness and maximization}. In what follows, we explain the underlying physics of how the counterflow depends on $G_\varphi, l, V_s,$ and $m$. From Fig. \ref{fig:robustness and maximization}(a), we observe that the counterflow measure is maximized at $G_{\varphi}^L/G_0=G_{\varphi}^R/G_0\approx 7$. At this optimal point, the spin triplet Cooper pair generation near the interface is high so that the Cooper pairs can carry a large spin supercurrent. At the same time, the pair breaking induced by the spin-mixing interfaces is small enough to not suppress the superconducting gap too much. At higher $G_{\varphi}$, $C$ decreases since superconductivity is suppressed, while at smaller $G_\varphi$, $C$ decreases since the triplet generation is suppressed. Exact compensation occurs for symmetric interfaces, while near-perfect counterflow compensation persists under substantial interface asymmetry. Numerically, the residual total spin current remains much smaller than the counterflow amplitude over most of the considered parameter range.

Figure \ref{fig:robustness and maximization}(b) shows that $C$ features non-monotonic behavior even as a function of length $l$. The counterflow vanishes in very short junctions, where superconductivity is suppressed, and approaches a finite value in the long-junction limit, where the two interfaces become independent. This cannot be attributed to the behavior of the superconducting gap, which increases monotonically with $l$. Instead, the physical mechanism can be understood as follows. The small maximum at intermediate length originates from the softened superconducting spectrum near the onset of superconductivity. In this regime, the interface-induced triplet correlations overlap throughout the superconductor, producing substantial subgap spectral weight. This enhances the dominant resistive contribution $M_{zz}\partial_yh_z$ within the low-energy window sampled by the spin distribution, and hence also the compensating spin-supercurrent contribution. As the junction length increases, the order parameter approaches its bulk value and the subgap spectral weight decreases. This reduces the counterflow toward its finite single-interface limit.

When calculating the $V_s$ and $m$ dependence of the counterflow, the superconducting state enters a bistable regime near the superconducting transition, where the Usadel equation has both normal and superconducting solutions corresponding to local minima of the free energy. We restrict the following discussion to the thermodynamically stable branch. The dependence of the counterflow on $V_s$ and $m$ is shown in Figs.~\ref{fig:robustness and maximization}(c) and (d), respectively. The two dependencies can be either monotonic or nonmonotonic, according to the value of the other parameter. A pronounced maximum appears only when the lower spin-split coherence peak of the density of states enters the nonequilibrium energy window, which at low temperature occurs approximately when $V_s \simeq \Delta-m$. Increasing $V_s$ extends the occupied energy window towards this peak, whereas increasing $m$ shifts the peak towards the occupied window. The non-monotonic behavior of the counterflow associated with the entry of this peak can then be understood physically as follows. When the spectral weight above the coherence peak in the spin-resolved density of states enters the nonequilibrium window, a strong redistribution among the kinetic-mode contributions to the quasiparticle current occurs. This is physically expected because 
both the number and the spin character of the quasiparticles participating in transport then change. The spin-dependent transport coefficients therefore vary rapidly in this energy range, changing how gradients of the different distribution modes contribute to the physical $z$-polarized current.

As a concrete example, consider the contribution to $J_z^{\rm qp}$ mediated by the energy mode $h_0$. This contribution changes abruptly from being negligible to carrying a significant part of the quasiparticle spin current once either $m$ or $V_s$ is large enough for the lower coherence peak to enter the nonequilibrium energy window. This happens since $h_0$  is weighted by different densities of states in the two spin sectors. Consequently, the same nonequilibrium occupation corresponds to unequal numbers of spin-up and spin-down quasiparticles and can therefore generate spin accumulation and spin transport~\cite{bobkova_jetp_15, silaev2015long,krishtop_prb_15}. This effect becomes particularly strong when the lower spin-split coherence peak enters the nonequilibrium window, since one spin sector then acquires a much larger spectral weight than the other. In the present calculations, the resulting $h_0$-mediated contribution flows opposite to the total counterflow and thereby assists its suppression beyond the maximum and thus non-monotonic behavior, both as a function of $m$ and $V_s$ (see the End Matter). The remaining kinetic-mode contributions are reorganized at the same spectral threshold. Finally, we note that although the prediction of the counterflow is here done for a diffusive transport regime, the mechanism behind it does not rely on impurity scattering. Therefore, we also expect supercurrent counterflow for ballistic transport, so long as the spin-splitting field in the superconductor and spin-active interfaces are present.

\textit{Experimental detection}---
The counterflow suggests two complementary electrical readout mechanisms. First, when the normal reservoirs are spin-Hall injectors, they may themselves act as detectors. Coupling to the superconductor modifies the spin accumulation and spin backflow in the normal metal, producing a reciprocal change in electrical current at fixed voltage, or in voltage at fixed current. This approach requires no additional detector, but a quantitative prediction requires solving the normal and superconducting kinetic equations self-consistently with two-sided spin-active boundary conditions. Interfacial spin mixing may also produce a normal-state background, so the counterflow would generally appear as a change in injector loading rather than necessarily as a strict on--off signal. Alternatively, the quasiparticle component may be accessed through a local inverse spin-Hall response in the superconducting channel. In the perturbative impurity spin-orbit regime, spin-to-charge conversion enters at first order in the characteristic strength $\gamma$ of the spin-orbit-dependent impurity potential, whereas spin-orbit relaxation enters at order $\gamma^2$~\cite{sinova_rmp_15, espedal2017spin}. Since the quasiparticle spin current induced by counterflow below $T_c$ occurs already at $O(\gamma^0)$, it produces a transverse voltage at $O(\gamma)$. By contrast, the spin-orbit relaxation generating a longitudinal quasiparticle spin current in the otherwise current-free normal state above $T_c$ would give $J_{z}^{\mathrm{qp}}=O(\gamma^2)$. The corresponding inverse spin-Hall background then appears only at $O(\gamma^3)$. This parametric separation makes the weak spin-orbit coupling assumed in our calculations, $l \ll l_\text{sf}$, a minimally perturbative probe of the counterflow. We show that the quasiparticle spin flow induced by the counterflow effect is indeed larger than the spin current induced by spin-orbit relaxation in the End Matter for experimentally realistic magnitudes of the spin-orbit scattering. Moreover, we estimate that a superconducting enhancement of the spin-to-charge conversion comparable to that reported in related spin-split-superconductor systems \cite{jeon_acsnano_20} raises the counterflow-induced voltage to the experimentally accessible range of several nanovolts. This makes the inverse spin-Hall response a promising electrical signature of the effect. The signal could be increased further by using a superconducting material with a larger normal-state spin-Hall angle than Al, provided that its spin-relaxation length remains sufficiently long for the counterflow to develop. A complete detector model must additionally include possible energy-Hall and other mode-dependent contributions to the transverse electrical response \cite{kamra2024inverse}.

\begin{acknowledgments}
\acknowledgments

This work was supported by the Research Council of Norway
through Grant No. 353894 and its Centres of Excellence
funding scheme Grant No. 262633 “QuSpin.” Support from
Sigma2 - the National Infrastructure for High Performance
Computing and Data Storage in Norway, project NN9577K, is
acknowledged. 

\textit{Data availability:}  The code and data used to prepare this manuscript are available
in Zenodo \cite{zenodo}.

\textit{Disclosure of Generative AI use:} During the preparation
of this manuscript, the authors used generative AI tools. Claude Opus 4.8 and 5.5 were used to optimize the numerical
scripts for solving the Usadel equation, and to discuss the interpretation of the results. Microsoft Copilot GPT 5.6 was used as an interactive brainstorming aid
for discussing physical interpretations of the numerical results. All models were used
 for style editing of portions of the text and the figures. 
\end{acknowledgments}

\bibliography{references}

\clearpage
\onecolumngrid
\begin{center}
\textbf{End Matter}
\end{center}
\twocolumngrid
\setcounter{equation}{0}
\setcounter{figure}{0}
\setcounter{table}{0}
\makeatletter
\renewcommand{\theequation}{A\arabic{equation}}
\renewcommand{\thefigure}{A\arabic{figure}}
\renewcommand{\thetable}{A\arabic{table}}

\appendix
\section{Decomposition of the quasiparticle current}
\begin{figure}[h!!!!!!!!!]
    \centering
    \includegraphics{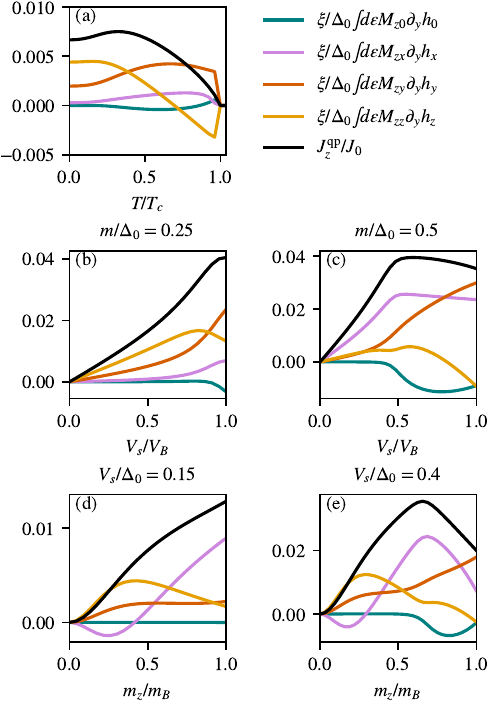}
    \caption{Decomposition of the quasiparticle current at $y=l$. The parameters used in (a) correspond to Fig. \ref{fig:mainres}(b). (b)-(c) correspond to Fig. \ref{fig:robustness and maximization}(c), and (d)-(e) correspond to Fig. \ref{fig:robustness and maximization}(d).}
    \label{fig:decomposition}
\end{figure}

The decomposition of the quasiparticle current into the four kinetic-mode contributions is shown in Fig. \ref{fig:decomposition}.

\section{Spin-orbit scattering and inverse spin Hall response}
\begin{figure}
    \centering
    \includegraphics{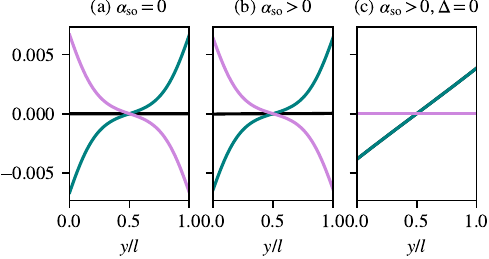}
    \caption{Spin currents in the presence of spin-orbit scattering. (a) Identical to Fig. \ref{fig:mainres}(a). (b) Including spin-orbit scattering $\alpha_\mathrm{so}/\Delta_0=0.005$. (c) The normal-state currents (no superconducting correlations).}
    \label{fig:spin-orbit}
\end{figure}

In the presence of spin-orbit scattering, an extra self-energy $\check{\Sigma}=i\alpha_\mathrm{so}\hat{\tau}_3\hat{\boldsymbol{\sigma}}\check{g}\hat{\tau}_3\hat{\boldsymbol{\sigma}}$ appears in the Usadel equation \cite{ouassou2017conservation, silaev2015long}. The spin-orbit scattering strength is $\alpha_\mathrm{so} = \hbar/(8\tau_\mathrm{so}) \propto \gamma^2$ \cite{sinova_rmp_15, espedal2017spin} where $\tau_\mathrm{so}$ is the characteristic relaxation time from spin-orbit scattering. Typical numbers for Al are $\tau_\mathrm{so} = \qty{45}{\ps}$~\cite{quay2018out, quay2015quasiparticle} and $\Delta_0=\qty{0.3}{\meV}$~\cite{court2008energy}, which gives $\alpha_\mathrm{so}/\Delta_0 \sim 0.005$. Figure \ref{fig:spin-orbit} shows how the spin-orbit scattering affects the spin currents. In Fig. \ref{fig:spin-orbit}(a), spin-orbit scattering is turned off, and the parameters used are the same as in Fig. \ref{fig:mainres}(a). With $\alpha_\mathrm{so}/\Delta_0=0.005$, the currents decrease slightly, as seen in Fig. \ref{fig:spin-orbit}(b). The total spin current $J_z$ is nonzero, but the spatial variation of $J_z$ is small compared to the counterflow currents $J_z^\mathrm{qp, super}$. In (c), the middle region is normal (non-superconducting). We see that spin-orbit scattering allows for a spatially varying quasiparticle current even in the absence of the counterflow decay channel, but the spatial variation of the current is smaller than in the counterflow regime. 

We first estimate the inverse spin-Hall induced electrical voltage in the normal-state $T>T_c$. The relation between the spin current and the transverse charge current is $J_Q = \theta_\mathrm{SH}\frac{2e}{\hbar}J_z^{\rm qp}$, where a typical normal-state spin-Hall angle $\theta_\mathrm{SH}$ for Al is of the order $10^{-4}$~\cite{valenzuela2006direct}. Using $N_0=\qty{2.4e28}{\eV^{-1}\m^{-3}}$ and $\xi=\qty{100}{\nm}$, we get $\theta_\mathrm{SH}\frac{2e}{\hbar}J_0  =  \qty{5.2e6}{\ampere \meter^{-2}}$. A typical value for $J_z^{\rm qp}/J_0$ induced by spin-orbit impurities in the normal-state $T>T_c$ in our results is $\sim 0.005$ [see Fig. \ref{fig:spin-orbit}], such that $J_Q \sim \qty{1e4}{\ampere \meter^{-2}}$.

The induced electric voltage is $V=\rho W J_Q$, where $\rho$ is the resistivity and $W$ the film thickness. With $\rho\approx 10^{-7}\qty{}{\ohm \m}$ \cite{valenzuela2006direct} and $W=50$ nm \cite{lopez_sst_25}, ensuring suppressed Meissner screening when going below $T_c$,
the voltage is $V \sim 0.1$ nV. This is a very small normal-state signal that is likely not detectable. In contrast, in the superconducting state, it has been predicted \cite{kamra2024inverse} and experimentally observed \cite{jeon_acsnano_20} that the spin Hall effect can be increased by several orders of magnitude. This means that one may expect a much larger signal induced by the counterflow current in the superconducting state compared to the value estimated above in the normal-state. A conservative enhancement of two orders of magnitude, smaller than the experimentally observed three orders of magnitude \cite{jeon_acsnano_20}, would lead to counterflow-induced inverse spin-Hall voltages of order $10~\mathrm{nV}$. Inverse spin-Hall voltages of such magnitude have previously been measured in Al~\cite{valenzuela2006direct}, demonstrating that the enhanced superconducting signal estimated here should be experimentally accessible. Its pronounced increase below $T_c$, together with the predicted dependencies on $T$, $m$, and $V_s$, would provide a characteristic electrical signature of the counterflow effect. 

\section{Maximal values of $V_s$ and $m$}
In Fig. \ref{fig:robustness and maximization}(c)-(d), the spin accumulation and the spin splitting are normalized on $V_B$ and $m_B$, respectively. Their values are given in Table \ref{tab:normalization bistable}.

\begin{table}[h!!!!!!!!!!!!]
    \centering
    \begin{tabular}{|c|c|}
    \hline
         $m=0.5\Delta_0$ & $V_B=0.91\Delta_0$    \\
    \hline
         $m=0.25\Delta_0$ & $V_B=0.69\Delta_0$   \\
    \hline
         $V_s=0.4\Delta_0$ & $m_B = 0.8\Delta_0$ \\
    \hline
         $V_s=0.15\Delta_0$ & $m_B = 0.58\Delta_0$ \\
        \hline
    \end{tabular}
    \caption{Normalization constants for Fig. \ref{fig:robustness and maximization}(c)-(d).}
    \label{tab:normalization bistable}
\end{table}

\end{document}